# Localized surface plasmon resonance in graphene nanomesh with Au nanostructures

Yang Wu,[1,2,3,†] Jing Niu,[1,†] Mohammad Danesh,[1] Jingbo Liu,[4] Yuanfu Chen,[4] Lin Ke,[3] Chengwei Qiu,[1] and Hyunsoo Yang[1,2,a)]

[1]Department of Electrical and Computer Engineering, National University of Singapore, 117576, Singapore
[2]Centre for Advanced 2D Materials, National University of Singapore, 117546, Singapore
[3]Institute of Materials Research and Engineering (IMRE), 3, Research Link, 117602, Singapore
[4]State Key Laboratory of Electronic Thin Films and Integrated Devices, University of Electronic Science and Technology of China, Chengdu 610054, China

[†]These authors contributed equally.
[a)] e-mail address: eleyang@nus.edu.sg

A hybrid structure of a graphene nanomesh with the gold nanodisks is studied to enhance the light absorption by localized surface plasmon resonance. From the reflection spectra of the visible range for graphene nanomesh samples without and with nanodisks, it is found that the absorption of graphene nanomesh structures is greatly enhanced in the presence of gold nanodisks around the resonance wavelength. Simulation results based on the finite-difference time-domain method support the experimental observations. This study demonstrates the potential of constructing graphene based photodetectors with a high light absorption efficiency and wavelength selectivity.

Graphene, with its unique electrical, optical, mechanical, and chemical properties, attracts enormous research interests since it was first exfoliated from bulk graphite in 2004.[1-4] Due to an ultrahigh carrier mobility and low momentum scattering rate, graphene electronics has been one of the most intensively investigated research topics.[5,6] It is believed that the potential of graphene in photonic and optoelectronics is also significant, as it exhibits both extraordinary electronic and optical properties.[3,7-9] Recently, graphene has been used as a saturable absorber in a mode-lock fiber laser.[10,11] A low sheet resistance as small as 30 ohm per square, a high transmittance of 97.7% over the entire visible range, and the strong endurance of graphene sheet makes graphene a suitable material as transparent electrodes.[12-14] In addition, graphene is considered as a promising material for photodetectors, since its optical absorption covers the visible as well as near infrared (NIR) range, which enables a broadband operation of the photodetectors.[14,15] A high carrier mobility of graphene allows the realization of ultrafast-response photodetectors.[16] Graphene devices are compatible with CMOS integration processes at wafer scales.[17] In a graphene photodetector with no source-drain biasing, a photo-response up to 40 GHz has been reported with a zero dark current.[16] Despite the above superior properties of graphene based photodetectors, there are two major drawbacks, such as a low photoresponsivity and low on-off ratio.

The low photoresponsivity is mainly due to the limited absorption of light by monolayer graphene, which is ~2.3%.[14,18] An efficient solution is to use localized surface plasmon resonances (LSPR),[19] and it has been shown that with plasmonic nanostructures,[9,20] the efficiency of graphene photodetectors is increased by 20 times.[21] In addition, by changing the dimension and period of nanodisks, the resonance can be adjusted to any wavelength in the visible range, which enables wavelength selectivity in graphene photodetectors.[9,21,22] The main reason of the low on-off ratio is due to the lack of a bandgap in graphene. Patterning graphene films into nanostructures is expected to solve the problem, because an energy gap can be formed through the quantum confinement effect of graphene in a two dimensional or zero dimensional structure.[23,24] So far, a variety of graphene nanostructures have been successfully fabricated, such as nanoribbons, nanodots, and nanomeshes. The optical characterization from a combined structure of graphene nanostructures with metal nanoparticles is an interesting research topic.[25,26] However, a hybrid structure with graphene nanomeshes which are filled up by plasmonic Au nanostructures has not been studied in photodetectors.

In this Letter, we propose a structure to enhance the light absorption by combining graphene nanomeshes and Au nanodisks. The devices are fabricated by filling the vacant holes of the graphene nanomesh with Au nanodisks. When LSPR is excited in metal nanodisks, the largest electric field is present among neighboring nanodisks which subsequently gives rise to an enhancement of light absorption. The experimental results demonstrate an enhancement of the absorption of incident light for the proposed graphene nanomesh with gold nanodisk structures. Furthermore, the experimental observations are confirmed by numerical finite-difference time-domain (FDTD) simulations.

Monolayer graphene is grown on copper foil by chemical vapor deposition (CVD) and then transferred on a 1 cm × 1 cm Si substrate capped with a 300 nm thick $SiO_2$ layer. In this study, tri-layer graphene is obtained by repeatedly transferring monolayer graphene film step by step for three times. Detailed growth condition and multi-step transfer process was similar to our previous report.[27] Electron beam lithography (EBL) is chosen to pattern graphene nanomesh. The dose of the EBL process is set as 80 μC/cm$^2$ with a 20 kV electron beam energy. Raman spectroscopy is utilized to monitor the quality of graphene before and after the sample preparation processes using a laser with the wavelength of 532 nm and the laser power of 0.5 mW is focused onto the samples. In the Raman measurements, the integration time is 10 s and each spectra is averaged for 5 times. The reflection spectra of the samples are obtained using a micro-spectrophotometer with a reflective objective lens.

The schematic of the device structure is shown in Fig. 1(a), where gold nanodisks are deposited in the holes of the graphene nanomesh. Nanodisks with different dimensions are fabricated; one has a 200 nm diameter with a 220 nm period and the other group has a 200 nm diameter with a 260 nm period. After the patterning, the device is scanned by atomic force microscopy (AFM) in a tapping mode at 340 kHz (probe type is TESPA) as shown in the inset of Fig. 1(a) and Fig. 1(b). Cylindrical graphene nanomeshes are formed before the deposition of gold nanodisks as shown in the inset of Fig. 1(a). After the deposition of gold, every empty cylinder is endowed with a hollow structure due to the shadowing effect during the deposition processes as shown in Fig. 1(b). The sample preparation processes are illustrated in Fig. 2, and the Raman spectroscopy is performed after each key process step. The Raman spectrum of bare CVD graphene (Fig. 2(a)) on the substrate is shown in Fig. 2(b). The clear 2D and G peaks without a D peak demonstrate a good quality of graphene films[28]. The ratio between the G and 2D peak confirms tri-layer graphene.

In the first step, a PMMA/MMA double layer electron beam resist is spin coated onto graphene. Then EBL is utilized to pattern the photoresist. Next, cylindrical graphene nanomesh is patterned by oxygen plasma (chamber pressure: 10 mTorr, etching time: 40 seconds, power: 20 Watt, and oxygen flow rate: 20 SCCM) with the patterned photoresist layer as an etch mask (Fig. 2(c)), and the Raman spectrum is repeated at this point as shown in Fig. 2(d). The inset of Fig. 2(d) shows the cross-section view of the nanomesh structure. After that, a 10 nm thick Au thin film is deposited by a thermal evaporator. Finally, the lift-off process removes the EBL photoresist and creates the pattern with nanodisks in the holes of the graphene nanomesh. The schematic of the final device is shown in Fig. 2(e), while the Raman spectrum for this structure is in Fig. 2(f). The inset of Fig. 2(f) is the cross-section view of the device.

Interestingly, a small red shift of the G peak is observed. From Lorentz fitting, the peak values are 1598.1 cm$^{-1}$, 1594.2 cm$^{-1}$, and 1592.0 cm$^{-1}$ in Fig. 2(b), 2(d). and 2(f), respectively, which can be due to the dopants induced Fermi level change of graphene. In addition, the D peaks are developed in Fig. 2(d) and 2(f), which indicates defects in graphene due to patterning. However, the full width of half maximum (FWHM) of 2D peaks is similar (31.7 cm$^{-1}$ for Fig. 2(b), 32.5 cm$^{-1}$ for Fig. 2(d), and 33.9 cm$^{-1}$ for Fig. 2(f)). In the Raman spectroscopy, the laser spot size (tens of microns) is much larger than the characteristic size (20 nm or 60 nm) of the graphene nanomesh, therefore, both the graphene films and etched graphene edges are examined during the Raman measurements. Therefore, it is reasonable to conclude that the D peak is mainly introduced by the edge of the holes in graphene nanomeshes, whereas graphene in-between these patterned holes is still in the early phase of amorphization,[29] which does not affect the functionality of graphene. The low intensity in the Raman signal is due to the small area of graphene after patterning.

In order to have a clear understanding of the role of the graphene nanomesh and Au nanodisks on the effect of LSPR, substrates without and with a graphene film are patterned during the nanomesh preparation process and the nanomesh structures are prepared without and with nanodisks. Therefore, total four types of samples are fabricated: a bare graphene thin film, patterned graphene nanomesh, nanodisk arrays without graphene, and nanodisk arrays with graphene nanomesh.

The optical characteristics are examined with an UV-Visible-NIR range micro-spectrophotometer (CRAIC) system. Measurements are conducted from four different sample structures, with the reference data from a Si/SiO$_2$ substrate. In order to evaluate the changes of the optical characteristics from the patterned samples, their contrast spectra with respect to that from the substrate and bare graphene are plotted in Fig. 3. The relative contrast is calculated using $C(\lambda) = (R_0(\lambda) - R(\lambda))/R_0(\lambda)$, where $R_0(\lambda)$ represents the reflection spectrum of either a Si/SiO$_2$ substrate or bare graphene, and $R(\lambda)$ is the reflection spectrum of the patterned structures. Figures 3(a) and 3(b) are the data from the samples with a 220 nm nanomesh period, while Figs. 3(c) and 3(d) are from a 260 nm nanomesh period. All the samples have a hole/nanodisk diameter of 200 nm. Figures 3(a) and 3(c) show the contrast spectra of four different structures with respect to a Si/SiO$_2$ substrate, whereas Figs. 3(b) and 3(d) show the contrast spectra with the bare graphene film as the contrast reference.

As shown in Fig. 3(a) the contrast spectrum of bare graphene film has the Fabry-Perot interference induced peaks at ~ 350 nm and 570 nm, which agrees well with a previous report.[30] For both Au nanodisk array and graphene nanomesh with Au nanodisks, the contrast is greatly enhanced at peak positions. In order to exclude the effect introduced by the intrinsic graphene property, the contrast spectra is plotted with respect to a bare graphene film as shown in Fig. 3(b). For the graphene nanomesh without Au nanodisks, two valleys are observed instead of peaks. This is because the patterning process etched graphene from the bare graphene film, therefore, the contrast of the graphene nanomesh is comparably less around the wavelength of the maximum contrast of the graphene film. On the other hands, the contrast of graphene nanomesh with Au nanodisks still shows two peaks. One of the peaks is ~ 610 nm, which agrees with the resonance wavelength of LSPR gold nanodisks with a diameter ~ 200 nm.[31] The other peak at ~ 360 nm could be due to the LSPR quadrupole resonance in the nanodisks.[32] The peaks in the contrast spectrum represent the smallest reflection, thus maximum absorption, as commonly known that LSPR enhances the light absorption. Consequently, the presence of Au nanodisks in the graphene nanomesh structure enhances the absorption of the structure. It is clear that the difference between the spectrum of only Au nanodisks and graphene nanomesh with Au nanodisks is introduced by the absorption of graphene nanomesh. Thus, the Au nanodisk array improves the absorption of graphene nanomesh compared to that of a bare graphene nanomesh. For different graphene nanomesh with a period of 260 nm in Fig. 3(c,d), the enhancement factor is even greater than that of a period of 220 nm in Fig. 3(a,b). An increase of the absorption loss in the 220 nm separation can be attributed to two effects. First, the relative area of the gold covering the 220 nm separation is higher than that of the 260 nm separation, resulting in a higher amount of loss to occur. The second reason is that the 220 nm separation has a smaller gap (~20 nm) in comparison to the 60 nm gaps in the 260 nm separation case. These smaller gaps create much stronger hybridized modes and light confinement, resulting in higher losses.

Using the FDTD method the structure was modelled with a 3D unit cell with periodicities of 220 nm in Fig. 4(a,b) and 260 nm in Fig. 4(c,d). The reflection spectra for five different configurations (Si/SiO$_2$, Si/SiO$_2$/graphene, Si/SiO$_2$/graphene nanomesh, Si/SiO$_2$/Au nanodisks, and Si/SiO$_2$/graphene nanomesh with Au nanodisks) were calculated, and the contrast ratios with respect to Si/SiO$_2$ and Si/SiO$_2$/graphene are plotted in Fig. 4. In both Fig. 4(a) and 4(c), the graphene nanomesh has a lower absorption than bare graphene films at the peak positions, similar to the experimental data in Fig. 3(a) and 3(c). Consequently, the calculated contrasts with respect to bare graphene show negative values for graphene nanomesh at the resonance wavelengths as shown in Fig. 4(b) and 4(d), in line with the experimental results in Figs. 3(b) and 3(d). The existence of Au nanodisks alone without graphene enhances the light absorption at the peak positions. However, the hybrid structure of Au nanodisks with graphene nanomeshes enhances the absorption more significantly, as shown in Fig. 4. Some dissimilarities are present between the simulation data and experimental results, which can be attributed to the defects in graphene films, imperfection in patterned shapes, unintentionally doping in graphene and some deviation in the simulation parameters.

The loss (absorption) mechanism in graphene and nanodisks is very different. In the Au nanodisk arrays, the loss is mainly attributed to the LSPR. On the other hand, in graphene, the loss is arising from its broadband optical absorption. Interestingly, we see that the loss mechanism in graphene is changed when the gold nanodisks are added on graphene. To elaborate, the contrast of the graphene nanomesh in comparison to the bare graphene film is negative at the peak positions. However, when the gold nanodisks are added to the graphene nanomesh, the total contrast actually increases in comparison to the case of Au nanodisks alone. This indicates that the Au nanodisks have effectively changed the loss mechanism in conjunction with the graphene nanomesh. The loss introduced by the graphene nanomesh is higher than the loss of a flat graphene layer. The Au nanodisks have LSPRs that highly confine electromagnetic energy in their surrounding environment. This confinement of electromagnetic energy around the Au nanodisks that cover graphene, results in a higher interaction between the incoming photons and the graphene sheet, causing the losses from the graphene sheet to increase. In other words, the Au plasmon resonances cause the light matter interaction in the graphene sheet to be enhanced. Thus, we attribute this loss mechanism to the LSPR enhanced absorption in graphene nano-structures.

In conclusion, graphene nanomesh structures without and with Au nanodisks are fabricated and studied. The Au nanodisks are utilized to fill the vacant part of the graphene nanomesh in order to maximize the light absorption. Both the experiment and simulation results show that Au nanodisks in the graphene nanomesh structure introduce an enhancement of light absorption at the LSPR wavelength. Therefore, the proposed hybrid structure with the graphene nanomesh and Au nanodisks can be considered as a promising structure to construct photodetectors with an enhanced absorption efficiency, which has wavelength selectivity depending on the size and material of nanostructures.

This research was supported by NRF-CRP "Novel 2D materials with tailored properties: beyond graphene" (No. R-144-000-295-281).

Figure caption

Fig. 1. (a) An illustration of the device structure with graphene nanomeshes and Au nanodisks. The inset is an AFM image for the graphene nanomesh without Au nanodisks. (b) An AFM image of patterned graphene nanomeshes with Au nanodisks.

Fig. 2. A schematic illustration of the sample prepration processes and Raman spectra of the samples. (a) Bare tri-layer graphene on a Si/SiO$_2$ substrate. (b) Raman Spectrum of tri-layer graphene. (c) A sample diagram after the graphene nanomesh is patterned by oxygen plasma. (d) Raman spectrum of the graphene nanomesh, and the inset is the cross-section view of the nanomesh. (e) Graphene nanomesh and Au nanodisk hybrid structure on a Si/SiO$_2$ substrate. (f) Raman spectrum of the device structure in (e), and the inset is the cross-section view of the sample in (e).

Fig. 3. Experimental results measured by a micro-spectrophotometer. (a) Relative contrast spectra with respect to a Si/SiO$_2$ substrate. The diameter of the nanomesh and nanodisks is 200 nm with a period of 220 nm. (b) Relative contrast spectra with respect to a bare graphene film from the data in (a). (c) Relative contrast spectra with respect to a Si/SiO$_2$ substrate. The diameter of the nanomesh and nanodisks is 200 nm with a period of 260 nm. (d) Relative contrast spectra with respect to a bare graphene film from the data in (c).

Fig. 4. Simulation data by FDTD method. Caculated relative contrast with respect to a Si/SiO$_2$ substrate with a period of 220 nm (a) and 260 nm (c). Simulated relative contrast with respect to a bare graphene film on Si/SiO$_2$ with a period of 220 nm (b) and 260 nm (d).

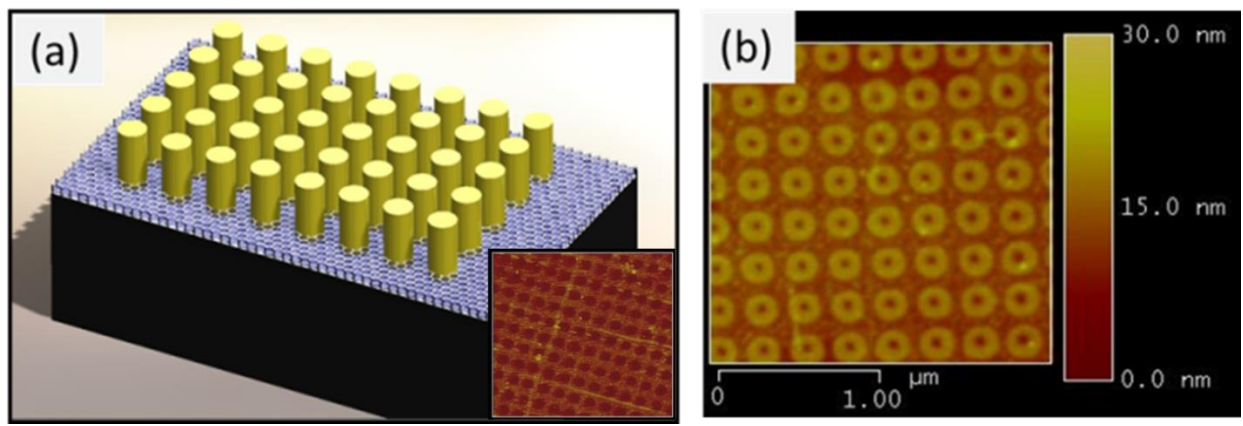

Figure 1

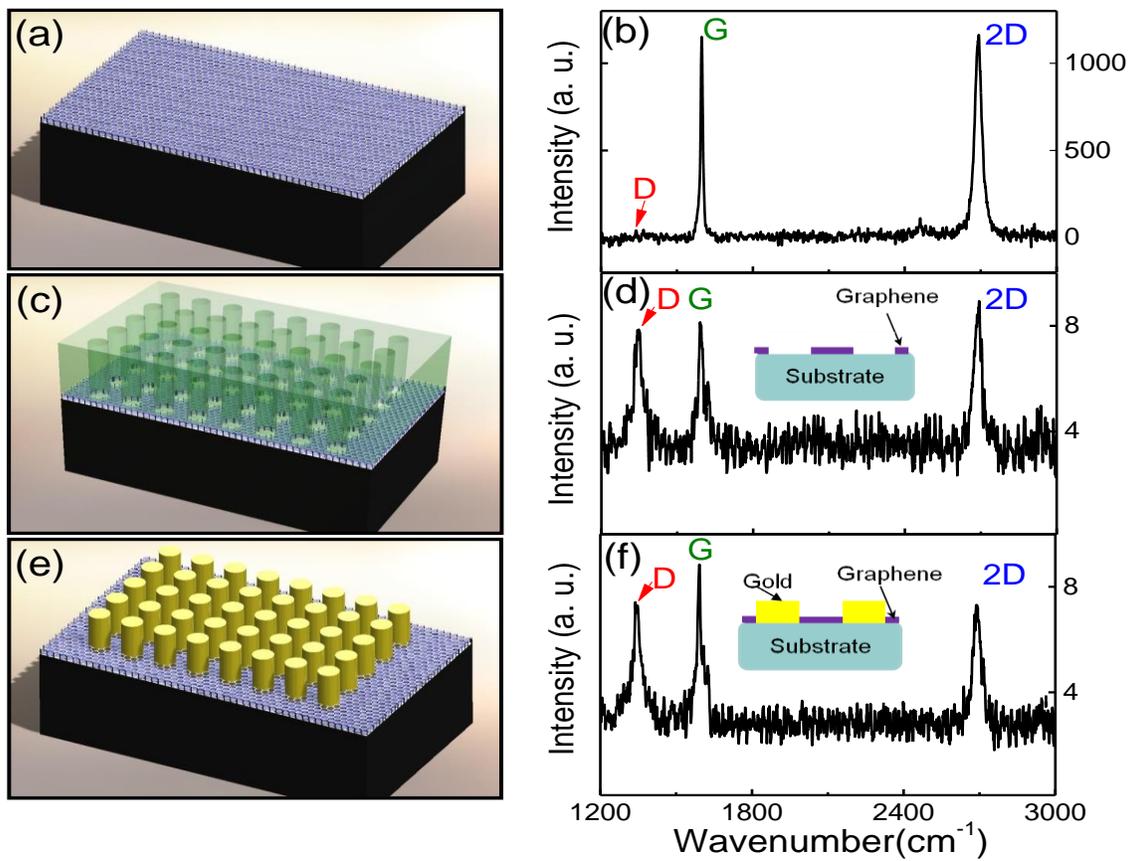

Figure 2

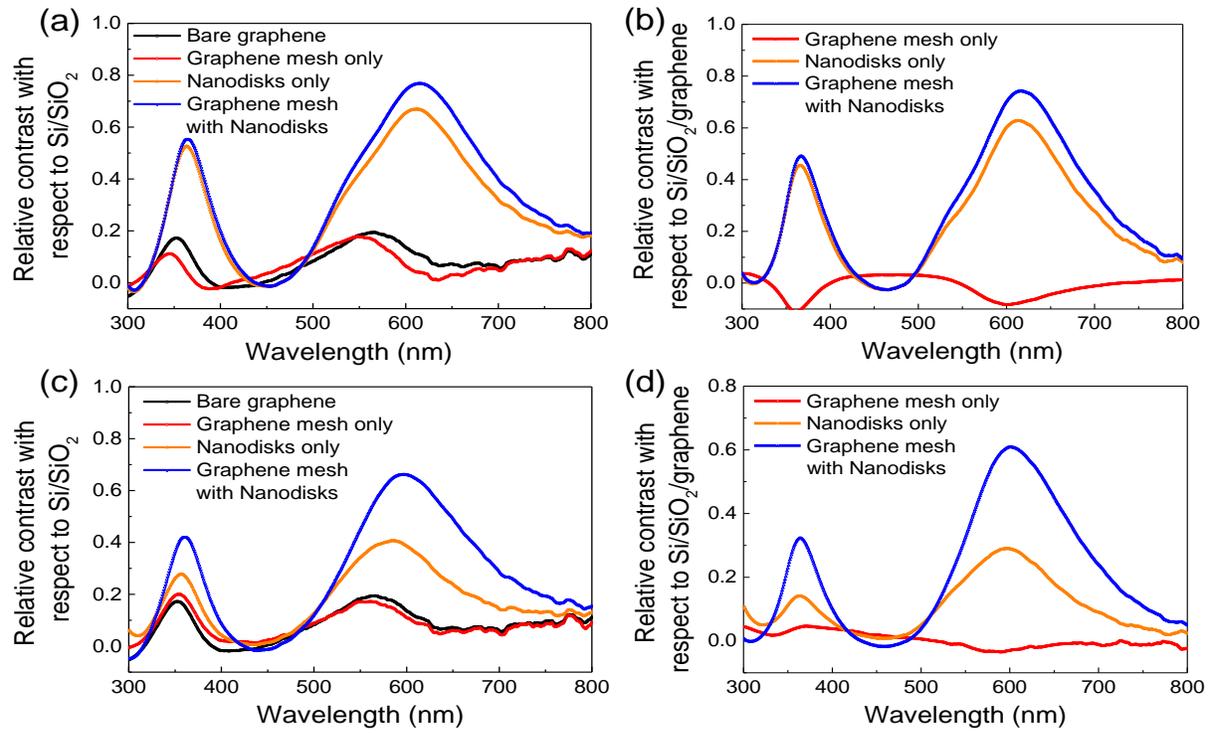

Figure 3

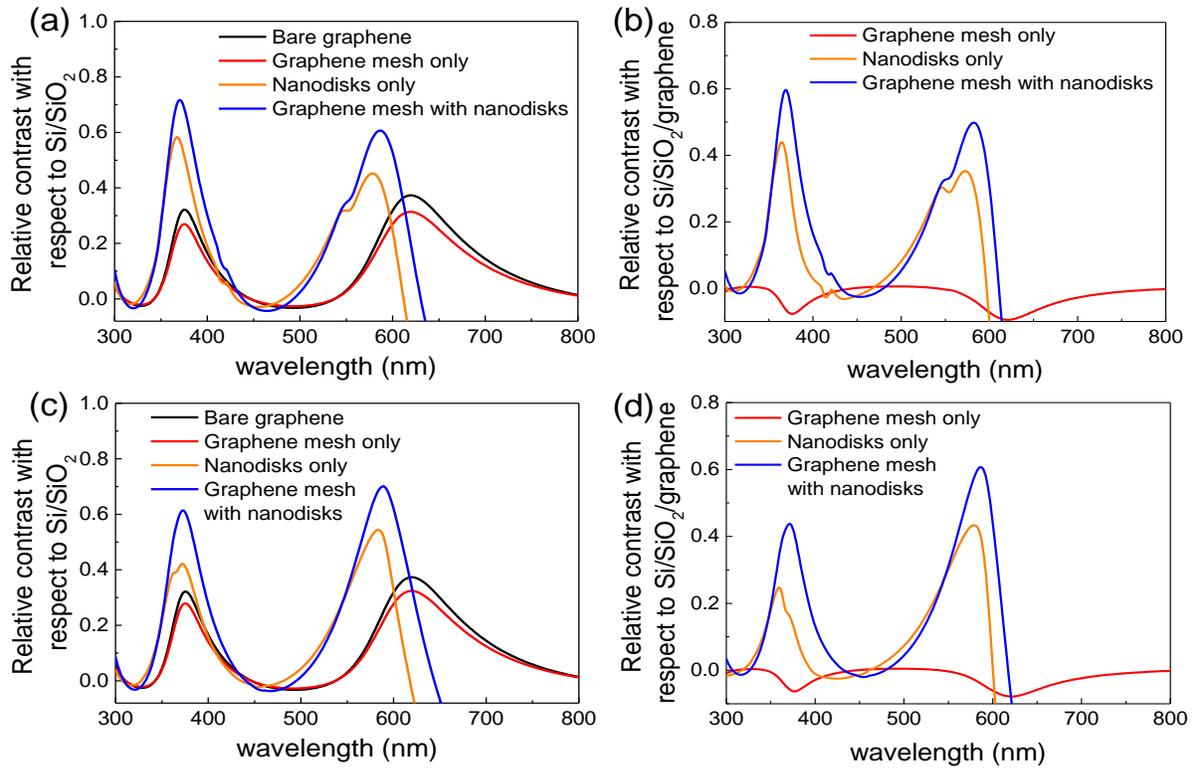

Figure 4